\documentclass[12pt,notitlepage,aps,amsmath,amssymb]{revtex4-1}

\usepackage{bm}
\usepackage{graphicx}

\usepackage{times}
\usepackage{mathptmx}

\newcommand{\ave}[1]{{\left<#1\right>}}

\newcommand{\nG}{\ensuremath{n_{\mathrm{G}}}}

\newcommand{\Eqref}[1]{Eq.~(\ref{#1})}

\newcommand{\Figref}[1]{Fig.~\ref{#1}}

\newcommand{\Ref}[1]{Ref.~\cite{#1}}

\newcommand{\etal}{{et al.}}
\newcommand{\gpi}{GPI}
\newcommand{\pdf}{PDF}
\newcommand{\sol}{SOL}

\newcommand{\nebar}{\ensuremath{\overline{n}_\mathrm{e}}}
\newcommand{\Ip}{\ensuremath{I_\mathrm{p}}}

\newcommand{\rmcm}{\ensuremath{\mathrm{cm}}}

\newcommand{\taud}{\ensuremath{\tau_\mathrm{d}}}
\newcommand{\tauw}{\ensuremath{\tau_\mathrm{w}}}
\newcommand{\Phirms}{\ensuremath{\Phi_\mathrm{rms}}}

\newcommand{\CJP}{{Czech.\ J.~Phys.}}
\newcommand{\JNM}{{J.~Nuclear Mater.}}
\newcommand{\NF}{{Nucl.\ Fusion}}

\newcommand{\PP}{{Phys.\ Plasmas}}

\newcommand{\PRL}{{Phys.\ Rev.\ Lett.}}
\newcommand{\PS}{{Phys.\ Scr.}}
\newcommand{\PPCF}{{Plasma Phys.\ Control.\ Fusion}}

\newcommand{\PCPS}{{Proc.\ Cambridge Phil.\ Soc.}}
\newcommand{\BSTJ}{{Bell Sys.\ Tech.\ J.}}
\newcommand{\PFR}{{Plasma Fusion Res.}}

\begin{document}

\title{Burst statistics in Alcator C-Mod SOL turbulence}

\author{O.~E.~Garcia$^{1,2}$}
\email{odd.erik.garcia@uit.no}
\author{I.~Cziegler$^{1,3}$}
\author{R.~Kube$^{1,2}$}
\author{B.~La{B}ombard$^2$}
\author{J.~L.~Terry$^2$}

\affiliation{$^1$ Department of Physics and Technology, University of Troms{\o}, N-9037 Troms{\o}, Norway}

\affiliation{$^2$ MIT Plasma Science and Fusion Center, Cambridge 02139, MA, USA}

\affiliation{$^3$ Center for Energy Research, University of California San Diego, CA, USA}

\begin{abstract}
Bursty fluctuations in the scrape-off layer (SOL) of Alcator C-Mod
have been analyzed using gas puff imaging data. This reveals many of the
same fluctuation properties as Langmuir probe measurements, including
normal distributed fluctuations in the near SOL region while the far SOL
plasma is dominated by large amplitude bursts due to radial motion of
blob-like structures.  Conditional averaging reveals burst wave forms with
a fast rise and slow decay and exponentially distributed waiting times.
Based on this, a stochastic model of burst dynamics is constructed.
The model predicts that fluctuation amplitudes should follow a Gamma
distribution. This is shown to be a good description of the gas puff
imaging data, validating this aspect of the model.
\end{abstract}

\maketitle

\section{Introduction}

Cross-field transport of particles and heat in the scrape-off layer (\sol)
of magnetically confined plasmas is dominated by radial motion of
blob-like structures \cite{review,garcia-tcv,labombard}. The average
cross-field particle and heat fluxes caused by such filaments depend on
their amplitude distribution and frequency of occurrence. The statistical
properties of plasma fluctuations in the \sol\ are thus crucial for
development of a first-principles physics based description of transport
and main-chamber interactions, and may also prove important for
understanding the empirical discharge density limit
\cite{review,garcia-tcv,labombard}.

Plasma fluctuations in the Alcator C-Mod \sol\ have been investigated
by analysis of data from gas puff imaging (\gpi) measurements at the
outboard mid-plane region in a set of ohmically heated, lower single
null discharges with a scan in line-averaged density. It is shown that
the \gpi\ diagnostic reveals many of the same fluctuation properties
as Langmuir probe measurements, including normal distributed fluctuations
in the near \sol\ region while the far \sol\ plasma with a broad particle
density profile is dominated by large amplitude bursts due to radially outward
motion of blobs \cite{review,garcia-tcv,labombard}. The fluctuation probability
distributions in the far \sol\ are strongly skewed and flattened with an
exponential tail towards large values for all line-averaged particle
densities \cite{review,garcia-tcv,labombard}.

The burst statistics are revealed by means of a standard conditional
averaging technique. \gpi\ measurements reproduce the familiar
asymmetric burst wave forms with a fast rise and a slow decay
and an average duration that is independent of the line-averaged
density \cite{garcia-tcv}. The waiting times between large amplitude
events are  found to be exponentially distributed for all line-averaged
densities. Based on these results, a stochastic model for the
intermittent fluctuations has been constructed. This model
reveals the importance of burst duration, waiting times and
amplitudes for large far \sol\ plasma densities and fluctuation
levels. From the model it is shown that the fluctuation amplitudes
follow a Gamma distribution and that there is a parabolic relationship
between the skewness and kurtosis moments. This compares favourably
with the \gpi\ measurements.

\section{Experimental setup}\label{sec:exp}

For this investigation we analyze \gpi\ data from a set of similar,
deuterium fuelled, ohmically heated, lower single null plasmas with
plasma current $800\,\text{kA}$ and toroidal magnetic field
$4.0\,\text{T}$. The \gpi\ diagnostic consists of a $9\times10$
array of in-vessel optical fibres with toroidally viewing, horizontal
lines of sight which are locally enhanced in the object plane by an
extended He gas puff from a nearby nozzle. Excitation of the He
neutral gas, and thus the intensity of the \gpi\ signals, is determined
by a combination of the local electron particle density and temperature 
\cite{cziegler}.  The fibres are coupled to high sensitivity  avalanche photo diodes
and the signals are digitized at a rate of $2\times10^6$ frames per second.
The viewing area covers the major radius from $88.00$ to $91.08\,\text{cm}$
and vertical coordinate from $-4.51$ to $-1.08\,\text{cm}$ with an
in-focus spot size of $3.7\,\text{mm}$ for each of the 90 individual
channels. The radial position of the last closed flux surface at the vertical
centre of the image, $Z=-2.61\,\text{cm}$, is in the range
from $89.4$ to $89.7\,\text{cm}$ for all the discharges presented here. The
limiter radius mapped to this vertical position is at $R=91.0\,\text{cm}$.
Further information about the \gpi\ diagnostic can be found in \Ref{cziegler}.

This paper reports on experiments performed during the FY2010 run
campaign using \gpi\ data from run 1100803. This comprises a
four-point scan in line-averaged density with the Greenwald fraction
$\nebar/\nG$ from $0.15$ to $0.30$. Here the  Greenwald density
is given by $\nG=\Ip/(\pi a^2)\,10^{20}\,\text{m}^{-3}$ where
$\Ip$ is the plasma current in units of MA and $a$ is the plasma minor
radius in units of meters. For the present density scan $\nG=5.26\times10^{20}\,\text{m}^{-3}$.
The condition at the outer divertor goes from sheath limited
at the lowest density to high recycling at the highest density in this scan.
For each discharge
the \gpi\ diagnostic yields $0.25\,\text{s}$ usable data time series during
the flat-top of the plasma current. By combining data from two discharges
at the same $\nebar$ and two nearby diode channels with identical
statistical properties, we obtain single-point time series of one second
duration which allows calculation of statistical averages with high accuracy.

\section{GPI measurements}

\begin{figure}
\includegraphics[width=7.5cm]{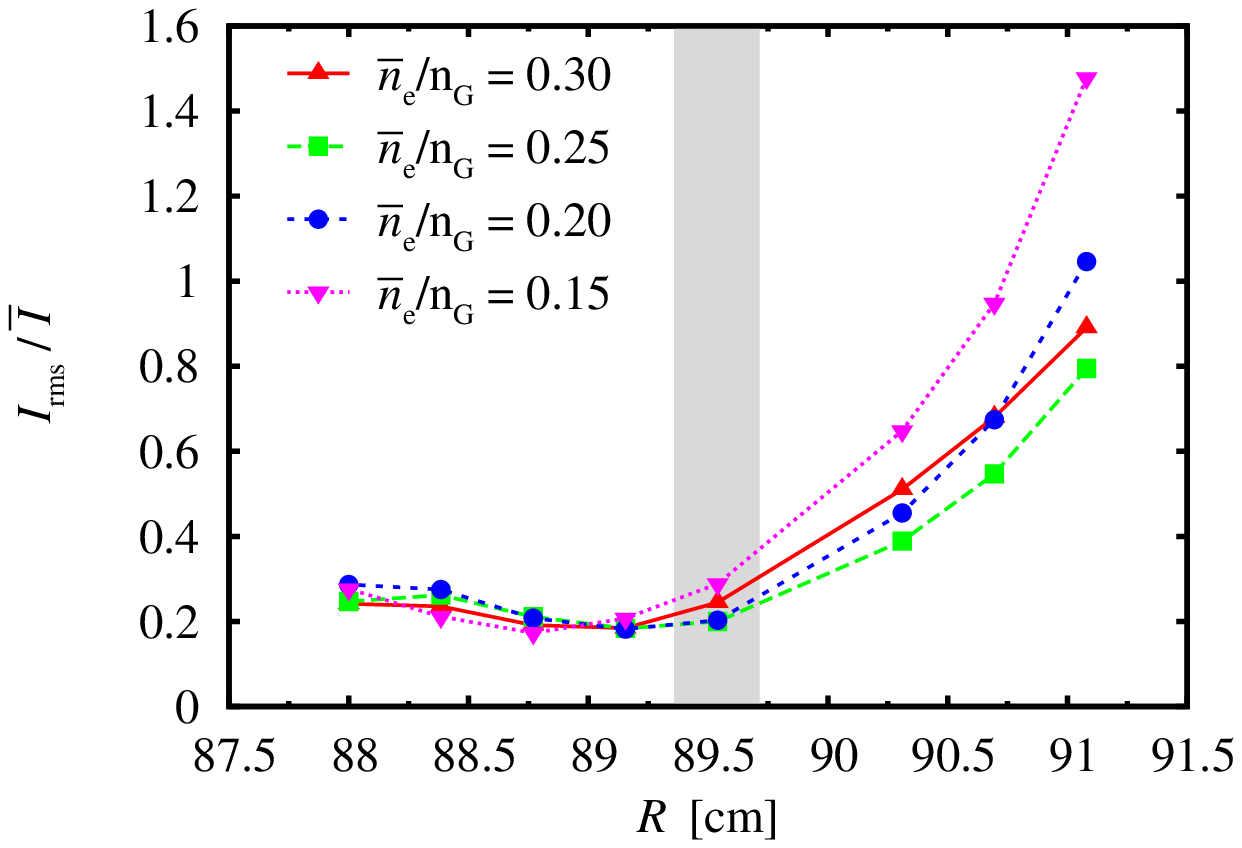}
\caption{Radial variation of relative \gpi\ intensity fluctuation level
 at $Z=-2.61\,\text{cm}$. The position of the last closed magnetic
flux surface is indicated by the shaded region.}
\label{fig:irfl}
\end{figure}

\begin{figure}
\includegraphics[width=7.5cm]{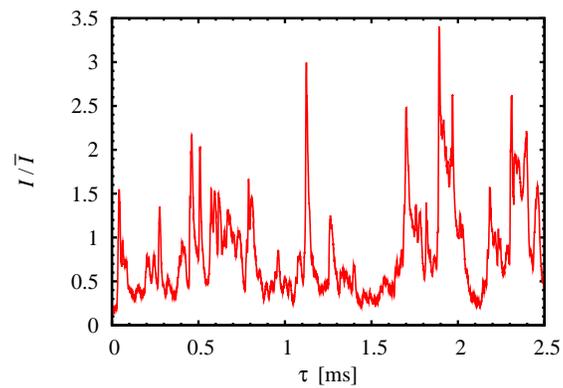}
\caption{Intensity signal for $\nebar/\nG=0.20$ at $(R,Z)=(90.69,-2.61)\,\rmcm$.}
\label{fig:iraw}
\end{figure}

\begin{figure}
\includegraphics[width=7.5cm]{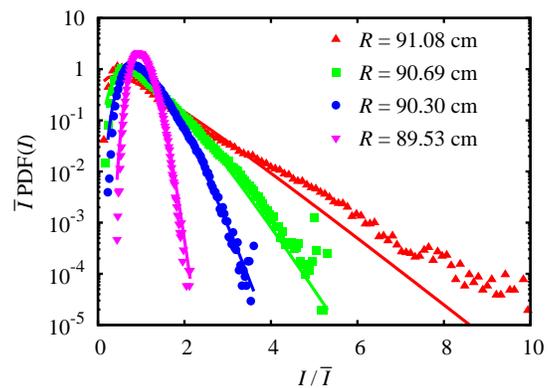}
\caption{\pdf\ of \gpi\ intensity signals for $\nebar/\nG=0.20$ at $Z=-2.61\,\rmcm$ and various radial positions in the \sol. The solid lines are
predictions of the model.}
\label{fig:ipdf}
\end{figure}

In \Figref{fig:irfl} we show the radial variation of the relative fluctuation
level of the \gpi\ intensity signals at $Z=-2.61\,\text{cm}$. The fluctuations
increase drastically in magnitude with radial distance into the \sol.
As shown in \Figref{fig:iraw},
the raw time series are here dominated by large-amplitude bursts
due to the radial motion of blob-like structures \cite{kube}. This results
in positively skewed and flattened probability density functions (\pdf s)
of the intensity signals. As an example of this, we present in \Figref{fig:ipdf}
the distribution function for the \gpi\ signals for $\nebar/\nG=0.20$ at
$Z=-2.61\,\text{cm}$ for the four \gpi\ fibre view positions radially outside
the separatrix. It is clearly seen that the \pdf\ changes from a
normal distribution in the near \sol\ region to strongly skewed
and flattened in the far \sol. At the limiter radius $R=91\,\text{cm}$,
the \pdf\ has an exponential tail towards large signal amplitudes. This is similar
to what has previously been found from Langmuir probe measurements
\cite{labombard,garcia-tcv}.

\begin{figure}
\includegraphics[width=7.5cm]{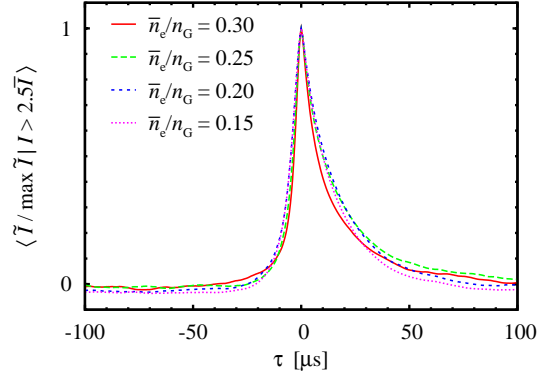}
\caption{Conditionally averaged wave form for \gpi\ intensity
burst amplitudes larger than $2.5$ times the mean value
at the reference position $(R,Z)=(90.69,-2.61)\,\text{cm}$.}
\label{fig:icav}
\end{figure}

\begin{figure}
\includegraphics[width=7.5cm]{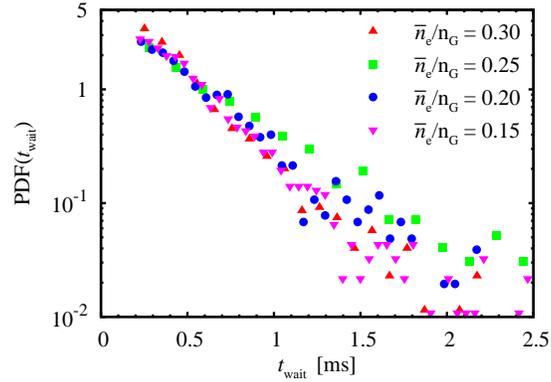}
\caption{\pdf\ of waiting times between burst events with maximum
amplitude larger than $2.5$ times the mean value at the reference
position $(R,Z)=(90.69,-2.61)\,\text{cm}$.}
\label{fig:wait}
\end{figure}

The radial motion of blob-like structures through the \sol\ results
in single-point recordings dominated by bursts with a fast rise and
a slow decay. This is demonstrated by the asymmetric wave form obtained
from conditional averaging presented in \Figref{fig:icav}. Here it is seen
that the average burst duration is the same for all line-averaged densities.
This is again similar to what has previously been found from probe
measurements as well as numerical turbulence simulations
\cite{garcia-tcv,garcia-esel}.
The waiting time between large amplitude bursts is also obtained
from the conditional averaging. As shown in \Figref{fig:wait}, the burst waiting
times are found to be exponentially distributed for all line-averaged
densities. An exponential distribution describes the time between
events for a Poisson process, in which events occur randomly and at
a constant average rate. The large-amplitude blobs appearing in the
far \sol\ are thus uncorrelated.

\section{Stochastic modelling}

The experimental measurements presented here suggest that
fluctuations in the \sol\ can be represented as a random sequence
of bursts similar to the classical ''shot noise" process \cite{campbell},
\begin{equation} \label{shotnoise}
\Phi(t) = \sum_k A_k\psi(t-t_k) ,
\end{equation}
where $t_k$ is the burst arrival time for event $k$ and $\psi$
is a fixed burst wave form. The burst waiting times, given by
$\tau_k=t_{k}-t_{k-1}$, are assumed to be exponentially distributed,
consistent with the results presented in \Figref{fig:wait}. The
individual burst wave forms will for simplicity be approximated
by a sharp rise followed by a slow exponential decay,
$\psi(t)=\Theta(t)\exp( -t/\taud)$, 
where $\Theta$ is the step function and the burst duration time
$\taud$ is taken to be constant. The burst amplitudes $A$ are
also taken to be exponentially distributed,
\begin{equation} \label{expamp}
P_A(A) = \frac{1}{\ave{A}}\,\exp\left( -\frac{A}{\ave{A}} \right) ,
\end{equation}
where $\ave{A}$ is the average burst amplitude. 
For the exponential burst wave form, the mean value
of the signal in \Eqref{shotnoise} is readily calculated
\cite{campbell,garcia-shotnoise},
\begin{equation} \label{phiave}
\ave{\Phi} = \frac{\taud}{\tauw}\,\ave{A} ,
\end{equation}
where $\tauw$ is the average burst waiting time. The above equation
elucidates the role of burst duration, waiting time and amplitude
for large \sol\ plasma densities.

In the case of exponentially distributed burst amplitudes,
the relative fluctuation level can be written as $\Phirms/\ave{\Phi}=(\tauw/\taud)^{1/2}$,
while the skewness and kurtosis moments for $\Phi$
are given by \cite{garcia-shotnoise}
\begin{equation} \label{KvsS}
S = \left( \frac{4\tauw}{\taud} \right)^{1/2} , \qquad
K = 3 + \frac{6\tauw}{\taud} .
\end{equation}
This shows that there is a parabolic relation between these moments given by $K=3+3S^2/2$. Like
the relative fluctuation level, also the skewness and kurtosis increases
with the ratio $\tauw/\taud$.  The parameter $\gamma=\taud/\tauw$
is thus a measure of intermittency in the signal given by \Eqref{shotnoise}.

\begin{figure}
\includegraphics[width=7.5cm]{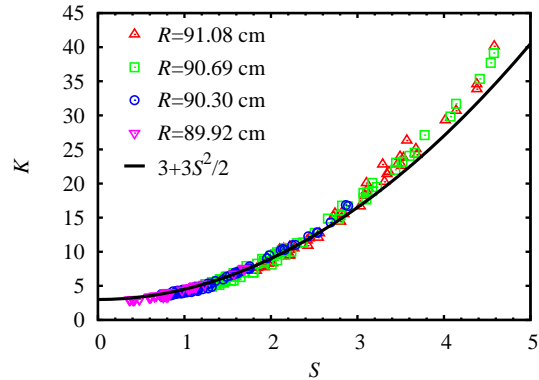}
\caption{Kurtosis versus skewness for the \gpi\ intensity signals in the
\sol\ for all discharges in the density scan. The full line is the prediction
of the model.}
\label{fig:kvss}
\end{figure}

In \Figref{fig:kvss} we present the kurtosis as function of skewness
calculated for all \gpi\ fibre view positions located in the \sol\ for the
discharges in the present density scan. It is clearly seen that the
skewness and kurtosis moments increase radially outwards in the
\sol, similar to the relative fluctuation level presented in \Figref{fig:irfl}.
Also shown in \Figref{fig:kvss} is the predicted parabolic relation
given by \Eqref{KvsS}. This is an excellent description of
the experimental data for all but the outermost diode channel
positions. Reduced emission from the neutral He gas is expected
for the cold and dilute plasma in the limiter shadow. Here
significant emission levels only arise
in the presence of large amplitude blob structures,  which leads
to excessively large higher order moments of the fluctuations.
This is likely the cause of the apparent discrepancy from a
parabolic relation for the signals from the outermost \gpi\ views in
\Figref{fig:kvss}. Such a parabolic relation between the skewness
and kurtosis moments has been found for many other plasma
experiments \cite{sattin,graves}.

For exponentially distributed burst waiting times
and amplitudes, the \pdf\ for $\Phi$ given by the model
\eqref{shotnoise} can be shown to be a Gamma distribution
\cite{garcia-shotnoise},
\begin{equation} \label{gamman}
\ave{\Phi}P_\Phi(\Phi) = \frac{\gamma}{\Gamma(\gamma)}
\left( \frac{\gamma\Phi}{\ave{\Phi}} \right)^{\gamma-1}
\exp\left( - \frac{\gamma\Phi}{\ave{\Phi}} \right) ,
\end{equation}
where the scale parameter is given by $\ave{\Phi}/\gamma$ and
the shape parameter is $\gamma=\ave{\Phi}^2/\Phirms^2$. In
\Figref{fig:ipdf} we have presented the corresponding Gamma
distribution, which is seen to be an excellent fit to the data for
the innermost \sol\ positions. Note that the \pdf s for the
experimental data comprise more than four decades on the
ordinate axis---a consequence of the long data time series
used in this analysis. In the limiter shadow the \pdf\
of the measured signals have a strongly elevated tail which is
likely due to suppressed emission in the cold far \sol\ plasma as
discussed above. The Gamma distribution has previously been
found to accurately describe the ion saturation current signal
in the \sol\ of TCV across a broad range of plasma parameters
\cite{graves}.

\section{Conclusions}

\gpi\ measurements in the \sol\ of Alcator C-Mod have shown that plasma
fluctuations are dominated by large amplitude bursts due to radial
motion of blob-like structures. The burst wave form is
asymmetric with a fast rise and slow decay. Conditional averaging
reveals that the burst waiting times and amplitudes are exponentially
distributed, thus blobs appearing in the far \sol\ are uncorrelated.
The \pdf\ of the \gpi\ intensity signal changes from a normal
distribution in the near \sol\ to strongly skewed and flattened with
an exponential tail in the far \sol. A stochastic model of burst dynamics
is constructed based on three parameters: burst duration, waiting time
and amplitude. Consistent with the predictions of the model,  the
\pdf s of the \gpi\ intensity fluctuations are well described
by a Gamma distribution and there is correspondingly a parabolic
relation between the skewness and flatness moments. These results
indicate that the stochastic model is an accurate description of the
burst dynamics in the tokamak \sol.

\clearpage

\end{document}